\documentclass[conference]{IEEEtran}
\IEEEoverridecommandlockouts

\usepackage{pbalance}

\IEEEoverridecommandlockouts 
\usepackage{subfig}
\usepackage{graphicx}
\usepackage{algorithmic}
\usepackage{algorithm}
\usepackage[OT4,T1]{fontenc}
\usepackage{amssymb}
\usepackage[cmex10]{amsmath}
\usepackage{amsthm}
\usepackage{url}
\usepackage{multirow}

\usepackage{orcidlink}
\usepackage{cleveref}
\usepackage{csquotes}

\usepackage{todonotes}

\newtheorem{remark}{Remark}[section]
\newtheorem{example}[remark]{Example}

\newcommand\copyrighttextmanual{%
	\footnotesize \textcopyright 2024 IEEE.  Personal use of this material is permitted. Permission from IEEE must be obtained for all other uses, in any current or future media, including reprinting/republishing this material for advertising or promotional purposes, creating new collective works, for resale or redistribution to servers or lists, or reuse of any copyrighted component of this work in other works.}
	\newcommand\copyrightnoticemanual{%
		\begin{tikzpicture}[remember picture,overlay]
			\node[anchor=south,yshift=10pt] at (current page.south) {\fbox{\parbox{\dimexpr\textwidth-\fboxsep-\fboxrule\relax}{\copyrighttextmanual}}};
		\end{tikzpicture}%
	}

\title{Comparing Lazy Constraint Selection Strategies in Train Routing with Moving Block Control}

\author{\IEEEauthorblockN{Stefan Engels\IEEEauthorrefmark{1}\,\orcidlink{0000-0002-0844-586X} and 
Robert Wille\IEEEauthorrefmark{1}\IEEEauthorrefmark{2}\,\orcidlink{0000-0002-4993-7860}}
\IEEEauthorblockA{\IEEEauthorrefmark{1}Chair for Design Automation, Technical University of Munich (TUM), 80333 Munich, Germany}
\IEEEauthorblockA{\IEEEauthorrefmark{2}Software Competence Center Hagenberg GmbH (SCCH), 4232 Hagenberg, Austria} \\
\IEEEauthorblockA{Email: \{stefan.engels, robert.wille\}@tum.de}
}

\crefname{table}{Tab.}{Tab.}
\crefname{equation}{Eq.}{Eq.}
\crefname{section}{Sec.}{Sec.}
\crefname{figure}{Fig.}{Fig.}
\crefname{example}{Ex.}{Ex.}
\crefname{observation}{Obs.}{Obs.}

\begin{document}
\maketitle              %
\copyrightnoticemanual

\begin{abstract}
Railroad transportation plays a vital role in the future of sustainable mobility.
Besides building new infrastructure, capacity can be improved by modern train control systems, e.g., based on moving blocks.
At the same time, there is only limited work on how to optimally route trains using the potential gained by these systems.
Recently, an initial approach for train routing with moving block control has been proposed to address this demand.
However, detailed evaluations on so-called \emph{lazy constraints} are missing, and no publicly available implementation exists.
In this work, we close this gap by providing an extended approach as well as a flexible open-source implementation that can use different solving strategies.
Using that, we experimentally evaluate what choices should be made when implementing a lazy constraint approach.
The corresponding implementation and benchmarks are publicly available as part of the \emph{Munich Train Control Toolkit}~(MTCT) at \url{https://github.com/cda-tum/mtct}.
\end{abstract}

\section{Introduction}
\IEEEPARstart{S}{ustainable} transportation systems are becoming increasingly important.
Because of this, the demand for railway transportation is constantly increasing.
Since building new tracks to increase capacity is resource- and time-consuming, train control systems should also be utilized to increase capacity.

Because trains cannot operate on sight due to their long braking distances, such systems are used to prevent collisions.
Most notable systems are the \emph{European Train Control System}~(ETCS), the \emph{Chinese Train Control System}~(CTCT), or the \emph{Positive Train Control}~(PTC) \cite{Pachl2021} as well as \emph{Communication Based Train Control}~(CBTC) for metro systems \cite{Schnieder2021a}.
While these systems differ in detail, the main concepts are very similar.
New specifications allow trains to follow each other more closely on existing infrastructure and at the same level of safety.
In the ideal case, trains can operate under a so-called moving block control, which provides enormous potential for increased capacity.

However, the most efficient specification does not help without methods to optimize train movements that use this potential.
Respective optimization tasks using classical (i.e., \enquote{old}) specifications are well studied \cite{Borndoerfer2018}.
At the same time, there is only a little work on routing under moving block control \cite{Schlechte2022,Klug2022}, none of which is available \mbox{open-source}.

Since the number of constraints preventing collisions is enormous and, at the same time, many of them are not needed to describe an optimal solution, a lazy approach is used.
First, the problem is optimized without these conditions.
During the solving process, violated constraints are iteratively added until a feasible (hence, optimal) solution is obtained.
There are different strategies on which (lazy) constraints to add in each iteration.
However, to the best of our knowledge, they have not previously been compared, and it is hard to do corresponding evaluations ourselves due to the lack of available implementations.

In this work, we aim to improve upon the aforementioned.
The resulting source code is included in the open-source \emph{Munich Train Control Toolkit}~(MTCT) available at \url{https://github.com/cda-tum/mtct}.
The solving strategy and other parameters can be chosen flexibly.
This allows for experimental evaluations, in which we analyze what strategy should be implemented using a lazy approach in train routing under moving block.
Additionally, the proposed model extends previous solutions to allow more general timetabling requests and can model train separation more precisely, especially in scenarios close to stations where a train might occupy multiple (short) train segments simultaneously.

The remainder of this work is structured as follows: \cref{sec:train-control} reviews the relevant principles of train control systems, \cref{sec:problem-contribution} describes the considered routing task and summarizes previous work as well as our contribution, and \cref{sec:milp-model,sec:lazy-headway} present the proposed approach(es). Finally, \cref{sec:experimental} contains an experimental evaluation, and \cref{sec:conclusions} concludes this paper.

\section{Train Control Principles}\label{sec:train-control}
Classically, a railway network is divided into fixed blocks.
Using \emph{Trackside Train Detection}~(TTD)~hardware, e.g., \emph{Axle Counters}~(AC), it is determined whether a particular block is occupied or not.
Because of this, the resulting blocks are also called TTD sections.
A following train can only enter a block once it is fully cleared by the previous train.

\begin{figure}[!t]
	\subfloat[Classical Block Signaling]{\includegraphics[width=\hsize]{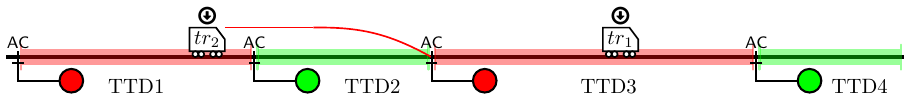}\label{fig:block-signaling}}
	\newline
	\vspace*{-1mm}
	\subfloat[Moving Block Signaling]{\includegraphics[width=\hsize]{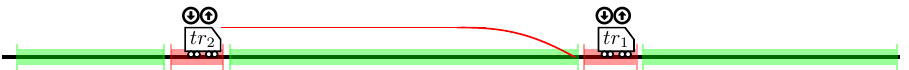}\label{fig:moving-block-signaling}}
	\vspace*{1mm}
	\caption{Schematic drawings of different signaling principles}
	\label{fig:etcs-levels}
	\vspace*{1mm}
\end{figure}

\begin{example}
	Consider two trains following each other on a single track as depicted in \cref{fig:block-signaling}. Train $tr_2$ can only move until the end of $TTD2$. It cannot enter $TTD3$ because it is still occupied and, hence, might have to slow down in order to be able to come to a full stop before entering the occupied block section. \label[example]{example:classical-signaling}
\end{example}

Modern control systems allow for more efficient headways.
A train equipped with \emph{Train Integrity Monitoring}~(TIM) can report its safe position to the control system.
Hence, no hardware is needed to safely separate trains.
Then, the network no longer has to be separated into fixed blocks.
In the best case, trains can follow at absolute braking distance.
Hence, shorter headway times are possible.
This so-called \emph{Moving~Block}~signaling has, e.g., been specified as an extension within ETCS Level 2 \cite{Siemens2023}, which is also formerly known as Level 3 \cite{Siemens2019}.
\begin{example}
	In contrast to \cref{example:classical-signaling}, consider a moving block control implemented in \cref{fig:moving-block-signaling}.
	Because trains operate at the ideal absolute braking distance, $tr_2$ can move up to the actual end of $tr_1$ (minus a little buffer).
	In particular, it can already enter what has been $TTD3$ previously.
	Hence, trains can follow each other more closely.
	\label[example]{example:moving-block}
\end{example}

\section{Problem Description and Contributions}\label{sec:problem-contribution}
In this work, we focus on moving block control systems.
This section briefly provides the problem description of a corresponding routing task, reviews the current state of the art, and motivates our work.

\subsection{Train Routing under Moving Block Control}\label{sec:routing-moving-block}
Train routing is the task of determining when and where trains are driving on the respective network, given timetabling constraints.
This includes the choice of specific tracks and corresponding timings.
More formally, it is defined using the following notation:
\begin{itemize}
	\item[$\mathcal{T}$:] A set of trains and its relevant properties (e.g., length, maximal speed, acceleration, braking curves).
	\item [$\mathcal{N}$:] A railway network including vertices $V$ and (directed) edges $E$ as described in \cite{EngelsDesignTasks2023Preprint}.
	\item[$\mathcal{S}$:] A set of stations, where each station $\mathcal{S} \ni S \subseteq E$ is a subset of edges of the network $\mathcal{N}$.
	\item[$\mathcal{D}^{(tr)}$:] A set of demands for every train $tr \in \mathcal{T}$ consisting of
	\begin{itemize}
		\item a weight $w^{(tr)} \geq 0$ of importance,
		\item an entry node $v_{in}^{(tr)} \in V$ together with a desired entry interval $[\underline{t}_{in}^{(tr)}, \overline{t}_{in}^{(tr)}]$,
		\item an exit node $v_{out}^{(tr)} \in V$ together with a desired exit interval $[\underline{t}_{out}^{(tr)}, \overline{t}_{out}^{(tr)}]$, as well as,
		\item a set of stations $S_i^{(tr)} \in \mathcal{S}$ together with
		\begin{itemize}
			\item an interval $[\underline{\alpha}_i^{(tr)},\overline{\alpha}_i^{(tr)}]$ in which the train should arrive at the station,
			\item an interval $[\underline{\delta}_i^{(tr)},\overline{\delta}_i^{(tr)}]$ in which the train should depart from the station, and
			\item a minimal stopping time $\Delta t_i^{(tr)} \geq 0$.
		\end{itemize}
	\end{itemize}
\end{itemize}
Having this notation, the goal is to determine an optimal routing.
In this setting, optimality is defined as minimizing the (weighted) exit times such that all schedule demands are obeyed and the constraints by a moving block control system are satisfied.

\subsection{State of the Art and Contributions}\label{sec:solution-approach}
Train routing and related timetabling tasks under classical train control have long been considered and are well-studied~\cite{Borndoerfer2018}.
On modern control systems using so-called hybrid train detection, routing is considered in algorithms to design optimal (virtual) section layouts by using SAT \cite{Wille2021}, A* \cite{Peham2022}, or \emph{Mixed Integer Linear Programming}~(MILP) \cite{Engels2023}.
While the arising questions are similar, these solutions do not utilize the full potential of moving block.

To the best of our knowledge, \cite{Schlechte2022} is the first approach that considers optimal routing of trains specifically under moving block control.
They describe a MILP formulation to solve a routing problem similar to the one considered in this paper.
Say $s$ describes the position and $t$ the time; one could say that their formulation models the function $t(s)$ at discrete positions given by vertices of the network.
Since trains cannot pass each other on a given edge, this seems to be a reasonable simplification while still being able to model at a decent level of accuracy.

However, the number of constraints to ensure that trains keep enough distance and do not crash into each other is rather big.
At the same time, most of these are unnecessary because trains operating at different network parts will not collide even without explicit constraints.
Because of this, one can first optimize without them.
If this yields a collision-free solution, the problem is solved.
Otherwise, constraints preventing collisions from arising need to be added during the solving process as so-called \emph{lazy constraints}.
By doing this, the same optimal solution is obtained; however, only a small number of the original constraints is considered.
This can be beneficial, especially for large models, as discussed in their follow-up work \cite{Klug2022}.

At the same time, this previous approach comes with some downsides:
\begin{itemize}
	\item Both trains and stations are single points without length.
	The authors claim this is not a problem because the length can be integrated as a buffer in the headway.
	However, especially in station environments, this might not be feasible.
	For example, some stations separate a platform into sections.
	A long train might occupy all of a platform, whereas two smaller trains can stop simultaneously (in different sections of the platform).
	Those scenarios cannot be modeled using the previous approach.
	\item There are different strategies to select which (lazy) constraints to add.
	This constitutes a trade-off: adding only a few lazy constraints in each iteration is quickly possible, however, many iterations might be needed until a collision-free solution is reported; on the other hand, adding many lazy constraints simultaneously increases the time spent in every iteration but, at the same time, likely reduces the number of iterations needed.
	In \cite{Klug2022}, no evaluation of selection strategies for lazy constraints is provided.
	\item The implementation of the solution is not publicly available.
	This prevents us from doing corresponding evaluations and restricts the proposed solution's applicability.
\end{itemize}

Overall, this motivates an alternative MILP formulation, which properly considers train separation even on shorter edges by considering the respective train lengths as well as incorporating more flexible timetable requests.
At the same time, we aim to shed light on which strategy for lazy constraint selection might be best by conducting a corresponding evaluation.
Finally, we provide a flexible \mbox{open-source} implementation at \url{https://github.com/cda-tum/mtct}, thus allowing the community to access such methods.

\section{MILP Model}\label{sec:milp-model}
This section presents the MILP model motivated in \cref{sec:solution-approach}.
For reasons of comprehensibility, we limit ourselves to the relevant variables and constraints.
The interested reader can find the complete model in the open-source implementation available at \url{https://github.com/cda-tum/mtct}.
\subsection{Symbolic Formulation}
\begin{figure}[t]
	\centering
	\includegraphics[width=\hsize]{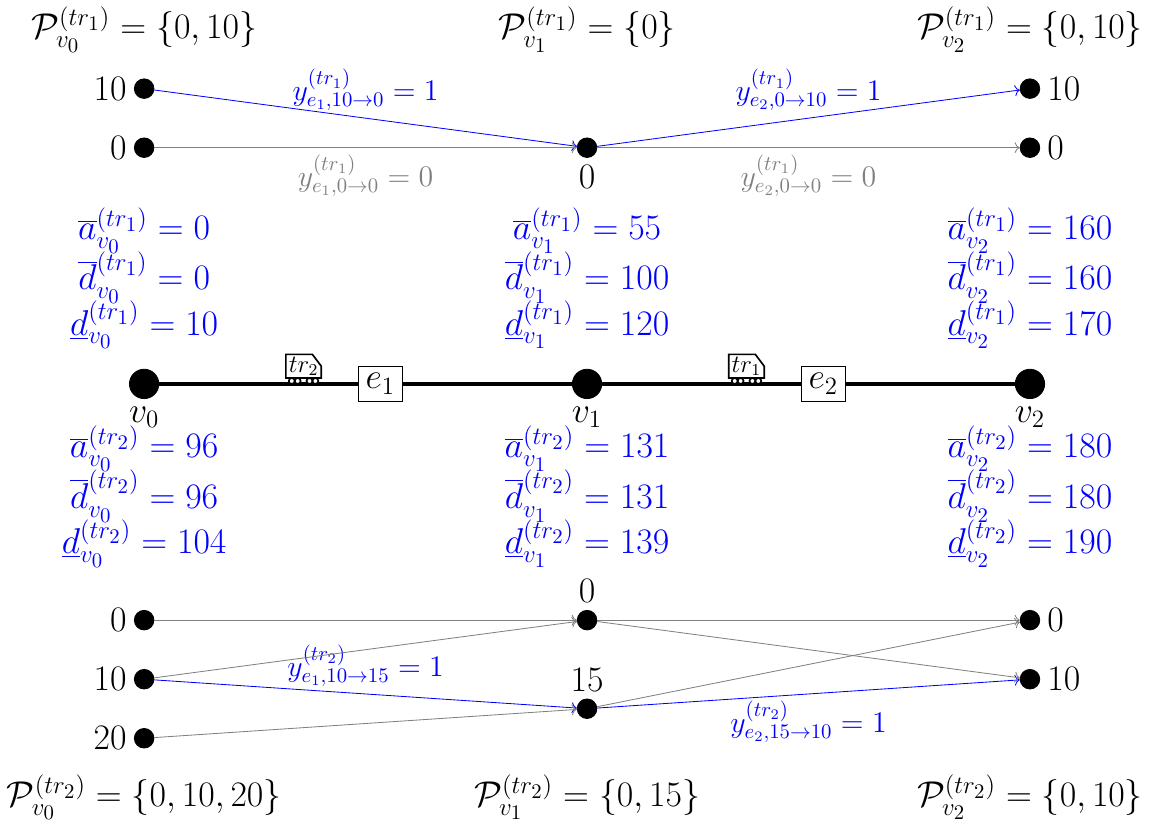}
	\caption{Example Setting for Symbolic Formulation}\label{example:symbolic-formulation}
\end{figure}

To model the approach, we need variables describing each train's routes and relevant timings.
As discussed in \cref{sec:solution-approach}, we follow the basic strategy by \cite{Schlechte2022} with slight extensions to incorporate the actual train lengths by tracking each train's rear point.
Hence, we include the following variables:
\begin{itemize}
	\item $x_e^{(tr)} \in \{0,1\}$ denotes whether a certain edge $e \in E$ is used by train $tr \in \mathcal{T}$.
	\item $\overline{a}_v^{(tr)} \in \left[0, \overline{t}_{out}^{(tr)}\right]$ is the time at which the front of train $tr \in \mathcal{T}$ arrives at $v \in V$.
	\item $\overline{d}_v^{(tr)} \in \left[0, \overline{t}_{out}^{(tr)}\right]$ is the time at which the front of train $tr \in \mathcal{T}$ departs from $v \in V$.
	\item $\underline{d}_v^{(tr)} \in \left[0, \overline{t}_{out}^{(tr)}\right]$ is the time at which the rear of train $tr \in \mathcal{T}$ departs from $v \in V$, hence, $tr$ has entirely left the previous edge.
\end{itemize}
The speed is included by extending the vertices accordingly.
Let $\mathcal{P}_v^{(tr)} \subseteq [0, v_{max}^{(tr)}]$ be a finite set of discretized velocities (paces) a train $tr \in \mathcal{T}$ might have at $v \in V$\footnote{For this work, we have used a uniform discretization of 10km/h as these values can be displayed by a standard speed indicator~\cite[Signal~Zs~3]{DIA2024}.}.
The extended graph has a (directed) edge $\epsilon_{e,p_1\rightarrow p_2}^{(tr)} \in E^{(tr)}$ connecting $(u,p_1)$ and $(v,p_1)$ for $e=(u,v) \in E$, $p_1 \in\mathcal{P}_u^{(tr)}$, and $p_2 \in\mathcal{P}_v^{(tr)}$ if, and only if, it is possible for train $tr \in \mathcal{T}$ to accelerate/decelerate from $p_1$ to $p_2$ while traveling on $e$.
For every such extended edge $\epsilon_{e,p_1\rightarrow p_2}^{(tr)}$, the variable
\begin{itemize}
	\item $y_{e,p_1\rightarrow p_2}^{(tr)} \in \{0, 1\}$ denotes whether train $tr \in \mathcal{T}$ uses the corresponding extended edge.
\end{itemize}
\begin{example}
	Consider again the setting of \cref{example:moving-block}, where two trains follow each other.
	For presentation purposes, we choose the segment to consist of three vertices.
	In \cref{example:symbolic-formulation}, the extended graph for $tr_1$ is drawn above and the one for $tr_2$ below the track.
	Values of relevant variables are written at the fitting places.
	Furthermore, note that $x_{e_1}^{(tr_1)} = x_{e_2}^{(tr_1)} = x_{e_1}^{(tr_2)} = x_{e_2}^{(tr_2)} = 1$.
\end{example}

\subsection{Constraints}
Of course, the above variables must be constrained to represent valid train movements.
In the following, we present the most important constraints.
We leave out some technical constraints needed to connect variables according to their desired purpose for ease of understanding.
Furthermore, we might formulate some constraints in its logical, rather than linear, form.
In either case, adding and reformulating the missing constraints using big-M and possibly some helper variables is straightforward.
They are irrelevant to presenting this work's crucial concepts and ideas.

\subsubsection{Valid Train Movements}
Each train $tr \in \mathcal{T}$ has to travel on a valid path from its entry to exit.
Hence,
\begin{equation}
	\sum_{e \in \delta^+\left(v_{in}^{(tr)}\right)} x_e^{(tr)} =
	\sum_{e \in \delta^-\left(v_{out}^{(tr)}\right)} x_e^{(tr)} =
	1
\end{equation}
where $\delta^+(\cdot)$ and $\delta^-(\cdot)$ denote outgoing and incoming edges respectively.
Moreover, a flow-conserving constraint (which also considers the respective velocity) has to be fulfilled at every other vertex.
Thus, for every \mbox{$v \in V - \left\{ v_{in}^{(tr)}, v_{out}^{(tr)} \right\}$} and $p \in \mathcal{P}_v^{(tr)}$ we have
\begin{equation}
	\sum_{\epsilon \in \delta^-(v,p)} y_\epsilon^{(tr)} = \sum_{\epsilon \in \delta^+(v,p)} y_\epsilon^{(tr)}
\end{equation}
within the velocity extended graph.
To prevent cycles, the in- and out-degrees are furthermore bound by one, respectively.

\subsubsection{Travel Times}
Denote by $\underline{\tau}^{(tr)}\left( \epsilon_{e,p_1\rightarrow p_2}^{(tr)} \right) \in \mathbb{R}_{\geq 0}$ the minimal time it takes train $tr \in \mathcal{T}$ to traverse $\epsilon_{e,p_1\rightarrow p_2}^{(tr)} \in E^{(tr)}$ of the velocity-extended graph.
For details on how these may be calculated, we refer to \cite[Fig. 3]{Schlechte2022}, but it suffices to consider them as an arbitrary oracle.
Analogously, let $\overline{\tau}^{(tr)}\left( \epsilon_{e,p_1\rightarrow p_2}^{(tr)} \right) \in \mathbb{R}_{\geq 0} \cup \{\infty\}$ denote the maximal time.
In this case, it is noted that a train might be allowed to stop on some of the edges, in which case $\infty$ is possible.
Again, we refer to \cite[Fig. 4]{Schlechte2022}.

Hence, assuming $e = (u,v)$, we have
\begin{align}
	\overline{a}_v^{(tr)} & \leq \overline{d}_u^{(v)} + \underline{\tau}^{(tr)}\left( \epsilon_{e,p_1\rightarrow p_2}^{(tr)} \right) + M \cdot \left(1 - y_{e,p_1\rightarrow p_2}^{(tr)} \right) \label{eqn:min-running-time} \\
	\overline{a}_v^{(tr)} & \geq \overline{d}_u^{(v)} + \overline{\tau}^{(tr)}\left( \epsilon_{e,p_1\rightarrow p_2}^{(tr)} \right) - M \cdot \left(1 - y_{e,p_1\rightarrow p_2}^{(tr)} \right) \label{eqn:max-running-time}
\end{align}
where $M \geq 0$ is large enough (e.g., $M = \overline{t}_{out}^{(tr)}$)\footnote{Note that \cref{eqn:max-running-time} is only added if $\overline{\tau}^{(tr)}\left( \epsilon_{e,p_1\rightarrow p_2}^{(tr)} \right) \leq \overline{t}_{out}^{(tr)}$ because otherwise bounding by the maximal travel time has no effect.} to ensure that the constraint is only activated if the respective edge is used.

Finally, 
\begin{equation}
	\overline{d}_v^{(tr)} \geq \overline{a}_v^{(tr)} \quad \forall v \in V, tr \in \mathcal{T} \label{eqn:positive-timing}
\end{equation}
and a train can only stop at a vertex (i.e., \enquote{$\neq$} in \cref{eqn:positive-timing}) if it has velocity 0, i.e.,
\begin{equation}
	\overline{d}_v^{(tr)} \leq \overline{a}_v^{(tr)} + M \cdot \sum_{\epsilon \in \delta(v,0)} y_\epsilon^{(tr)}.
\end{equation}
\subsubsection{Track Release}\label{sec:track-release}
In contrast to \cite{Schlechte2022}, we do not model a train as a single point.
This allows for more accurate train separation in the model.
For this, we need to relate the end of a train to its front.
Let $\mathcal{R} = \{e_1, \dots, e_k\} \subseteq E$ be a route starting in $v$, such that $\sum_{i=1}^{k-1} l(e_i) < l(tr) \leq \sum_{i=1}^k l(e_i)$\footnote{In general we denote by $l(\cdot)$ the length of an object}.
Similarly to above, let $\underline{\tau}_{\lambda \rightarrow \mu}^{(tr)}(\epsilon)$ and $\overline{\tau}_{\lambda \leadsto \mu}^{(tr)}(\epsilon)$ be the minimal and maximal travel time from point $\lambda$ to $\mu$ on the velocity extended edge $\epsilon \in E^{(tr)}$, where $0 \leq \lambda \leq \mu \leq l(\epsilon)$.
Then, the following bounds have to hold
\begin{align}
	x_e^{(tr)} = 1 \forall e \in \mathcal{R} & \Rightarrow \underline{d}_{u_1}^{(tr)} \geq \overline{d}_{u_k}^{(tr)} + \sum_{\epsilon \in \mathcal{E}_k} y_\epsilon^{(tr)} \cdot \underline{\tau}_{0 \rightarrow s}^{(tr)}(\epsilon) \label{eqn:track-release-1} \\
		x_e^{(tr)} = 1 \forall e \in \mathcal{R} & \Rightarrow \underline{d}_{u_1}^{(tr)} \geq \overline{a}_{v_k}^{(tr)} - \sum_{\epsilon \in \mathcal{E}_k} y_\epsilon^{(tr)} \cdot \overline{\tau}_{s \rightarrow l_k}^{(tr)}(\epsilon) \label{eqn:track-release-2}
\end{align}
assuming $e_i = (u_i, v_i)$, $s := l(tr) - \sum_{i=1}^{k-1} l(e_i)$, $l_k := l(e_k)$, and $\mathcal{E}_k$ being the set of all edges connecting $u_k$ to $v_k$ in the velocity extended graph.

While we chose to write the logical form in \cref{eqn:track-release-1,eqn:track-release-2} for better readability, they can easily be reformulated into linear constraints using big-M.
We do not add upper bounds because the objective of small headways pushes the variables down wherever needed.

\subsubsection{Headway}\label{sec:milp-headway}
Reference \cite{Schlechte2022} models train headways on single edges, which is precise if edges are rather long.
However, the braking distance considered might range multiple edges, particularly close to stations.
We use and proceed similarly to \cref{sec:track-release} to model this more precisely.
However, the length of the train is replaced by its braking distance.

On each edge $e \in E$, we introduce binary variables $o_e^{tr_1 \succ tr_2} \in \{0,1\}$, which is 1 if, and only if, $tr_1 \in \mathcal{T}$ follows $tr_2 \in \mathcal{T}$ on edge $e$.
The respective headway constraints relating $\overline{a}$ of the following and $\underline{d}$ of the preceding train are then analog to \cref{eqn:track-release-1,eqn:track-release-2}; however, with the additional conditions that the following train has a specific velocity and the respective ordering variable is one.

Similarly, one can proceed with trains traveling in opposite directions.
Then, however, the respective track segments behave like a TTD section, and a train's moving authority can only enter a track segment once the opposing train has entirely left it.

\subsubsection{Timetable}
Of course, also the timetable demands $\mathcal{D}^{(tr)}$ have to be satisfied.
Reference \cite{Schlechte2022} can bind the respective timing variables directly since the exact stopping points are predefined.
While, in our case, this is true for the entry and exit nodes, each stop at station $S_i^{(tr)} \in \mathcal{S}$ could be at a particular set of vertices, say $V_{S_i}^{(tr)}$.
For every such $v \in V_{S_i}^{(tr)}$, we add a respective binary variable $stop_{i,v}^{(tr)} \in \{0,1\}$.
Then,
\begin{align}
	stop_{i,v}^{(tr)} = 1 & \Rightarrow \overline{a}_v^{(tr)} \in [\underline{\alpha}_i^{(tr)},\overline{\alpha}_i^{(tr)}], \\
	stop_{i,v}^{(tr)} = 1 & \Rightarrow \overline{d}_v^{(tr)} \in [\underline{\delta}_i^{(tr)},\overline{\delta}_i^{(tr)}], \text{ and } \\
	stop_{i,v}^{(tr)} = 1 & \Rightarrow \overline{d}_v^{(tr)} - \overline{a}_v^{(tr)} \geq \Delta t_i^{(tr)}.
\end{align}
Again, these logical constraints can easily be reformulated into linear constraints using big-M.

\subsection{Objective}
Finally, the goal is to enable every train to leave the network as early as possible.
If a train leaves after its predefined earliest departure time, it is caused by the routing choice, not the respective request.
We minimize this difference according to the given weights, which we normalize to one.
Thus, the objective is given by
\begin{equation}
	\min \frac{1}{\sum_{tr \in \mathcal{T}} w^{(tr)}} \cdot \sum_{tr \in \mathcal{T}} w^{(tr)} \cdot \left( \underline{d}_{v_{out}}^{(tr)} - \underline{t}_{out}^{(tr)} \right).
\end{equation}

\begin{figure*}[!t]
	\centering
	\includegraphics[width=\hsize]{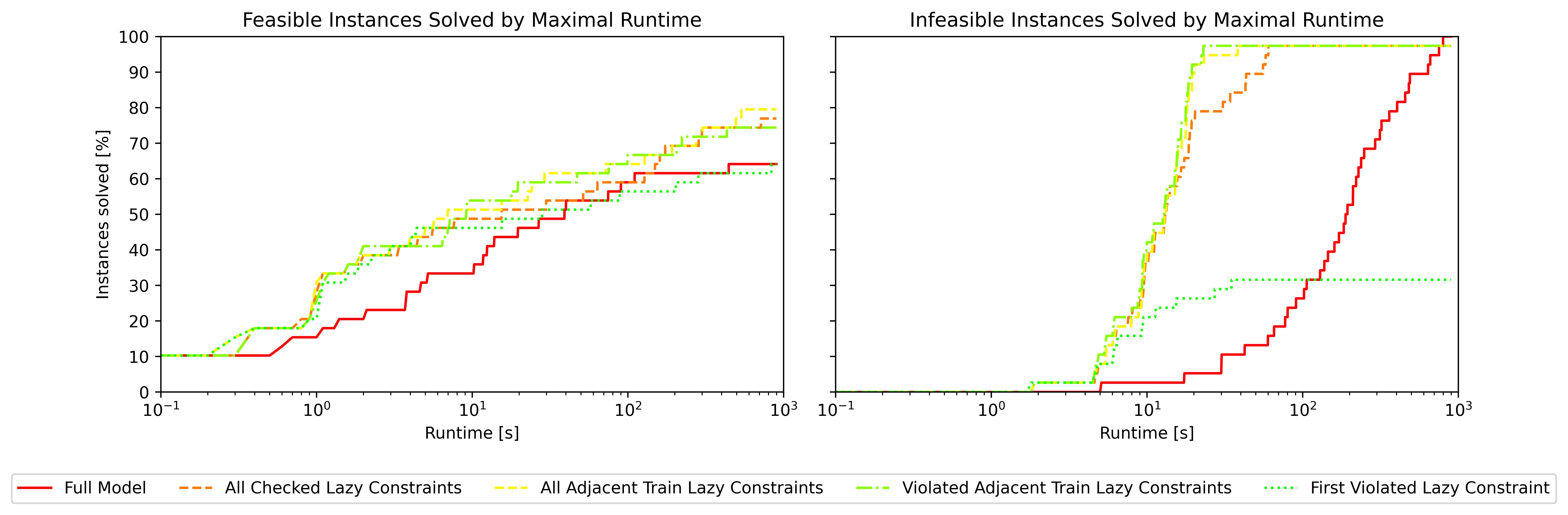}
	\caption{Runtime of Different Lazy Constraint Strategies}
	\label{fig:results}
\end{figure*}
\section{Lazy Headway Constraints}\label{sec:lazy-headway}

Note that there are many headway constraints of the form described in \cref{sec:milp-headway},
more precisely of order $\mathcal{O}\left( |\mathcal{T}| \cdot \sum_{tr \in \mathcal{T}} \sum_{v \in V} |\mathcal{P}_v^{(tr)}|\right) = \mathcal{O}\left( |\mathcal{T}|^2 \cdot |V| \cdot \overline{|\mathcal{P}|} \right)$, where $\overline{|\mathcal{P}|}$ denotes the average number of velocity extensions.
For instances with many trains on more extensive networks, the time to explicitly add all these constraints to a model is substantial.
However, most of these constraints are not explicitly needed because they describe a scenario far from optimal.
This motivates a lazy approach.

For this, we optimize using all except the headway constraints.
The obtained solution could violate some of the requirements.
If so, one has to add a set of violated constraints to the model and reoptimize, whereby the solver can use information from the previous iteration to warm start.
This procedure is continued until the solution is feasible and, hence, optimal.

However, the question arises of which constraints to add in each iteration.
A given conjectured solution determines each train's route and velocity profile uniquely.
Hence, at most $\mathcal{O}\left(|\mathcal{T}| \cdot \sum_{tr \in \mathcal{T}} |\mathcal{X}^{(tr)}| \right) =  \mathcal{O}\left(|\mathcal{T}|^2 \cdot \overline{|\mathcal{X}|}\right)$ conditions need to be checked, where $|\mathcal{X}^{(tr)}|$ denotes the number of edges used by a respective train and, again, $\overline{|\mathcal{X}|}$ the respective average.
In fact, the number is even smaller because only trains whose routes intersect need to be compared.

Note that the headway constraints are transitive in the sense that if $tr_1$ follows $tr_2$ at a safe distance at a given point, then it also safely follows all trains that might have passed before $tr_2$.
In particular, it suffices to check (and possibly add) only headways regarding the immediate preceding train.
This reduces the number of checked constraints to $\mathcal{O}\left(|\mathcal{T}| \cdot \overline{|\mathcal{X}|}\right)$.

Finally, one could either add all checked constraints (to give as much information as possible to the solver) or only such constraints that are violated (to not overload the solver with unnecessary information). In the extreme case, one could even only add one violated constraint and stop checking further constraints immediately.
In the first case, one is likely to only need a few iterations.
Conversely, a first violated constraint might be found very quickly; however, more iterations are needed in the end.
The best strategy might depend on the specific problem and is worth evaluating.

\section{Implementation and Evaluation}\label{sec:experimental}
The approach presented above has been implemented, made publicly available as \mbox{open-source}, and used to evaluate lazy constraint selection strategies.
It is included in the \emph{Munich Train Control Toolkit} available at \url{https://github.com/cda-tum/mtct} and under active development.
In this section, we describe both the resulting implementation as well as the evaluations and results obtained.
\subsection{Implementation}
We implemented the model described in \cref{sec:milp-model} using the C++ API of Gurobi \cite{GurobiOptimization2023}.
The resulting tool allows the user to choose between different strategies for lazy constraints to be added in each iteration by controlling various parameters.
Lazy constraints are implemented using the (custom) callback framework provided by Gurobi.

In the current version, the tool allows to e.g., compare the following selection strategies:
\begin{itemize}
	\item \enquote{\emph{Full Model}:} The entire model is explicitly constructed in advance and passed to the solver. No callback is used.
	\item \enquote{\emph{All Checked Lazy Constraints}:} In case of infeasibility, all $\mathcal{O}\left(|\mathcal{T}|^2 \cdot \overline{|\mathcal{X}|}\right)$ constraints corresponding to overlapping routes are added in each iteration regardless if they are violated or not. 
	\item \enquote{\emph{All Adjacent Train Constraints}:} Similarly, all checked constraints are added in case of infeasibility. However, only $\mathcal{O}\left(|\mathcal{T}| \cdot \overline{|\mathcal{X}|}\right)$ conditions corresponding to adjacent trains directly following each other are considered. 
	\item \enquote{\emph{Adjacent Violated Constraints}:} Again, only constraints corresponding to adjacent trains are considered. This time, however, only violated constraints are passed to the solver in each iteration. Conditions already fulfilled are ignored but might be added to a later callback.
	\item \enquote{\emph{Only First Violation}:} As soon as one violated constraint is found, only this one is added, and the callback is immediately aborted without checking the remaining conditions.
\end{itemize}

\subsection{Evaluation}\label{sec:experimental-evaluation}
We tested these different strategies on an Intel(R) Xeon(R) W-1370P system using a 3.60GHz CPU (8 cores) and 128GB RAM running Ubuntu 20.04 and Gurobi version 11.0.2.
As benchmarks, we use the railway networks and schedules from~\cite[Appendix A]{Engels2023}.
Additionally, we create random timetables of up to 50 trains on two of the networks, including the Munich S-Bahn Stammstrecke.
Since optimizing up to the millisecond is unreasonable, we stop at a proven optimality gap of 10 seconds.

The results are provided in \cref{fig:results}. 
On the x-axis, we plot the runtimes in seconds.
Note that we chose a logarithmic scale for better readability.
The y-axis provides the fraction of samples that were solved in the given time or faster.
The lines are monotonously increasing by design.
Generally, if a line is over/left of another line, the corresponding algorithm performs faster/better.

To avoid distorting the analysis because of infeasible instances, we present two plots.
On the left, we included instances known to be feasible; on the right, we included instances proven to be infeasible.
In the latter case, the time plotted corresponds to the time it took the proposed approach to prove infeasibility.

Clearly, the numbers confirm that a lazy approach is beneficial: one should not explicitly specify the entire model in advance.
The only exception to this is that only adding one constraint at a time performs even worse, which becomes especially clear when considering infeasible examples.

Among the other strategies, no one clearly outperforms the others.
At the same time, there seems to be a slight advantage of only considering adjacent trains directly following each other instead of all possible pairs of trains.
However, it is questionable if this effect is significant.

Overall, it seems reasonable to only add violated constraints corresponding to adjacent trains.
However, other strategies might also be beneficial depending on the context in which the algorithm is used since the observed benefit is only minor.

Having the proposed approach available as \mbox{open-source} will allow adding and evaluating further strategies easily.

\section{Conclusions}\label{sec:conclusions}
In this work, we considered train routing within a moving block environment.
We introduced a MILP formulation that can more accurately (than existing solution methods) model train separation on layouts with short track segments by incorporating the actual train length.
Moreover, we discussed how a lazy constraint approach can be implemented using different strategies in each callback.
Various such strategies have been implemented \mbox{open-source} and are available at \url{https://github.com/cda-tum/mtct}.
The user can control the parameters affecting the solving process.

An experimental evaluation confirms that the solution process benefits from the lazy approach as long as multiple constraints are added simultaneously.
On the other hand, there seems to be no significant difference between some of the tested strategies.
At the same time, the open source implementation allows for the use of different strategies depending on the instance, and it is not necessary to decide on the one and only best approach in this setting.

Previous work focuses on the optimal design of other modern train control systems relying on so-called hybrid train detection.
These systems combine the efficiency of moving block with the practicability of classical train control \cite{Bartholomeus2018}.
Design automation methods in this context must both route trains and place so-called virtual subsections.
However, both of these tasks alone are already hard, and optimization methods in this context can highly benefit if routing is considered separately \cite{Engels2023}.
While the details are out of scope for this paper, it is reasonable to believe that optimal routes under moving block are good choices in this case.
In particular, we aim to include this work (and possible future work on routing under moving block control) as a first step within an optimization pipeline for automated planning of train control systems with hybrid train detection, which, again, will also be made available open-source as part of the aforementioned Munich Train Control Toolkit.

\bibliographystyle{IEEEtran}
\bibliography{lit_etcs_combined}

\end{document}